\documentclass[%
 reprint,
 amsmath,amssymb,
 aps,
 prd
]{revtex4-1}

\usepackage[utf8]{inputenc}
\usepackage{graphicx}
\usepackage{dcolumn}
\usepackage{bm}
\usepackage{hyperref}

\newcommand{\expect}[1]{\langle #1 \rangle}

\begin{document}

\title{Gravitational collapse of thin shell of dust in shape dynamics}

\author{Furkan Semih Dündar}
\affiliation{Physics Department, Boğaziçi University, 34342, Istanbul, Turkey}
\email{furkan.dundar@boun.edu.tr}

\author{Zahra Mirzaiyan}
\affiliation{Physics Department, Isfahan University of Technology, 84156-83111, Isfahan, Iran}
\email{z.mirzaiyan@ph.iut.ac.ir}

\author{Metin Arik}
\affiliation{Physics Department, Boğaziçi University, 34342, Istanbul, Turkey}
\email{metin.arik@boun.edu.tr}

\begin{abstract}
We studied the gravitational collapse of a shell of dust in shape dynamics. We found out static and oscillatory solutions. In the large momentum limit we found out that the shell never reaches the singularity when the momentum of the shell is much larger than the mass of the shell in magnitude. The shell does not reach to the origin in a finite amount of time however when the momentum of the shell becomes comparable to minus the mass of shell, the large momentum approximation breaks down. Therefore more detailed future works hopefully may be able to answer the question of singularity formation in this setup.
\end{abstract}

\maketitle


\section{Introduction}

A re-interpretation of Einstein's general theory of gravitation (GR) \cite{gr-found} (see refs. \cite{gr-carroll,gr-wald} for a detailed account of the theory) named as shape dynamics (SD) \cite{sd-1,sd-found-1,sd-found-2} (see ref. \cite{tutorial} for a review) does not include local Lorentz symmetry and instead involves local scale (Weyl) symmetry. Historically, the development of SD comes after Julian Barbour's interpretation of Mach's principle \cite{mach}. Later, Barbour gives a completely relational account of $N$ particles with gravitational interaction in ref. \cite{barbour}. Although the foundations of SD were laid in refs. \cite{sd-found-1,sd-found-2,sd-1}, the form of SD as we understand it today has been found in refs. \cite{link1,link2}. What is basically done in \cite{link1,link2} is to remove the local Lorentz symmetry in GR and replace it with local scale symmetry.

In ref.~\cite{thinshell} the collapse of a thin shell of spherically symmetric dust under gravity has been investigated and it has been found out that the infalling matter does not cross the event horizon in a finite amount of asymptotic time, however, nevertheless leaves the Schwarzschild solution \cite{sch-german,sch-english} outside the shell. This is in agreement with the results from GR. In GR, an observer well outside the arena cannot observe matter crossing the event horizon but only approaching it in the $t \to \infty$ limit. Ref.~\cite{thinshell-compact} collapse of a single and two thin shells of spherical symmetry is studied. In the single shell scenario, they found out that the shell is static. This is mainly because there the space has the topology of $S^3$ (In our case it is $[0,1] \times S^2$). We did not study two shell scenario, so we leave it readers to read more about it through ref.~\cite{thinshell-compact}. 

\begin{figure}[ht]
	\includegraphics[width=0.2\textwidth]{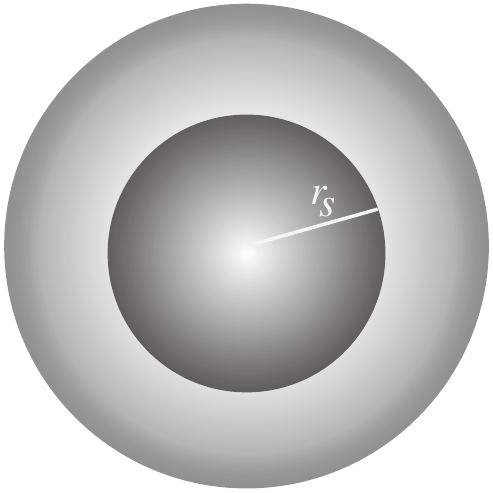}
    \caption{Ilustration of the configuration of the system. The dust shell is located with spherical symmetry at radius $r_s$.}
    \label{fig:rs}
\end{figure}

The solution of ref. \cite{thinshell} has downsides (as noted in their article) in the way that SD requires a compact space, whereas they considered events happening in an asymptotically flat space that is non-compact. We remedy this problem by considering a spherical shell of dust of radius $r_s$ in a 3-ball ($S_b \equiv [0,1] \times S^2$) with $r_s < 1$ and with boundary condition $(1,\theta,\phi) = (1,\pi-\theta,-\phi)$. See Figure~\ref{fig:rs}. Hence $S_b$ is compact without boundary where shape dynamics is well defined \cite{fate}.

The paper is organized as follows. In Section~\ref{sec:vacuum} the equations of motion in the vacuum are given for SD. In Section~\ref{sec:coupling} we couple a spherically symmetric thin shell of dust to shape dynamics and set up the jump conditions for the metric components. We investigate the time evolution of the shell in Section~\ref{sec:evolution}. Since the equations of motion are very complex to be solved analytically we solve them in various limiting cases and, finally, Section~\ref{sec:conclusion} is devoted to conclusions.

\section{The Vacuum Setup}\label{sec:vacuum}

We consider a spherically symmetric compact space, which is $[0,1] \times S^2$, with antipodal matching. In that space, we consider a spherically symmetric thin shell of dust located at coordinate radius $r_s$. Inside and outside $r_s$ the space is empty. For that purpose we need to describe metric and metric momenta in the vacuum. The most general spherically symmetric metric and metric momenta are as follows \cite{fate}:

\begin{align}
	g_{ab} &= \text{diag}(\mu^2,\sigma,\sigma \sin^2\theta),\label{eq:gab}\\
    p^{ab} &= \text{diag}\left(\frac{f}{\mu},\frac{s}{2},\frac{s}{2\sin^2\theta}\right)\sin\theta\label{eq:pab},
\end{align}

where $\xi^a = (\xi(r),0,0)$ is the shift vector field and $\mu,\sigma,f,s$ are functions of the radial coordinate $r$. In the CMC (Constant Mean extrinsic Curvature) gauge, the vacuum SD constraints are as follows \cite{tutorial}:

\begin{align}
	\mathcal{H} &= \frac{1}{\sqrt{g}}\left(p^{ab}p_{ab}-\frac{1}{2}p^2\right) - \sqrt{g}R,\label{eq:H}\\
    \mathcal{H}_a &= -2 \nabla_b p^b_a,\label{eq:Ha}\\
    p &= \sqrt{g}\expect{p},\label{eq:cmc}
\end{align}

where $\mathcal{H}$ is the ADM Hamiltonian, $\mathcal{H}_a$ is the diffeomorphism constraint and $p$ is the trace of the conjugate metric momentum. When the values of $g_{ab}$ and $p^{ab}$ is put in equations (\ref{eq:H}), (\ref{eq:Ha}) and (\ref{eq:cmc}) we get the following equations \cite{tutorial}:

\begin{align}
	& -\frac{1}{6\sigma\mu^2}\Big[\sigma^2 \mu s^2 + 4 f^2 \mu^3 - 4 f \sigma \mu^2 s + 12 \sigma \mu \sigma''\nonumber\\
    &\quad -12 \sigma \sigma' \mu' - 3\mu (\sigma')^2 - 12\sigma \mu^3 - \expect{p}^2 \sigma^2\mu^3\Big]=0,\label{eq:const1}\\
    & \mu f' - \frac{1}{2}s\sigma' = 0,\label{eq:const2}\\
    & \mu f + s\sigma - \expect{p}\mu\sigma = 0,\label{eq:const3}
\end{align}

where ($'$) stands for differentiation with respect to $r$. Using equations (\ref{eq:Ha}) and (\ref{eq:cmc})  gives us:

\begin{equation}
	\frac{\mu}{\sqrt{\sigma}}\left(f\sqrt{\sigma}-\frac{1}{3}\expect{p}\sigma^{3/2}\right)' = 0.
\end{equation}

Hence we obtain the solution as:

\begin{equation}
	f\sqrt{\sigma}-\frac{1}{3}\expect{p}\sigma^{3/2} = A(\tau),\label{eq:solf}
\end{equation}

where $\tau$ is York time. The Hamiltonian constraint can be written as follows \cite{tutorial}:

\begin{align}
	&-\frac{\sigma^{1/2}\mu}{\sigma'}\frac{\partial}{\partial r}\Big[\left(\frac{\sigma'}{\sigma^{1/4}\mu}\right)^2-4\sqrt{\sigma}-\frac{f^2}{\sqrt{\sigma}}\Big]\\
    &= \frac{2f\mu}{\sigma'}\left(f'+\frac{f\sigma'}{2\sigma}-\frac{\expect{p}\sigma'}{2}\right).\label{eq:H2}
\end{align}

The term inside the parenthesis on the right hand side of equation (\ref{eq:H2}) vanishes \cite{tutorial}, hence the term whose radial derivative is taken must equal to a constant \cite{tutorial}:

\begin{equation}
	\left(\frac{\sigma'}{\sigma^{1/4}\mu}\right)^2-4\sqrt{\sigma}-\frac{f^2}{\sqrt{\sigma}} = -8m.\label{eq:8m}
\end{equation}

When the solution (\ref{eq:solf}) is put into equation (\ref{eq:8m}) we obtain the $\mu^2$ as:

\begin{multline}
	\mu^2 = (\sigma')^2\Big(A^2\sigma^{-1} + (\frac{2}{3}\expect{p}A - 8m)\sigma^{1/2}\\
     +4\sigma + \frac{1}{9}\expect{p}^2\sigma^2\Big)^{-1}\label{eq:mu2}
\end{multline}

We have found $\mu^2$ as in (\ref{eq:mu2}) however we know that it should be greater than zero because the three dimensional metric should be regular. This condition puts a restriction on the radial derivative of $\sigma$ ($\sigma'$) and the $\sigma$ itself. This issue is handled in \cite{tutorial} in detail and we suggest readers read this note.

\section{Coupling a  Thin Shell of Dust to Gravity and the Jump Conditions}\label{sec:coupling}

The problem of coupling a thin shell of dust to shape dynamics is done in \cite{thinshell}. We will follow its definitions to settle a ground for discussion. We suppose there are $n$ particles each of mass $m_0$. The ADM constraints for a single particle is as follows \cite{thinshell}:

\begin{align}
	\mathcal{H} &= \delta(x^a-y^a) \sqrt{g^{ab}p_a p_b + m_0^2},\\
    \mathcal{H}_a &= \delta(x^a-y^a) p_a.
\end{align}

If we take the continuum limit of particles on a shell of radius $r_s$, the constraints become \cite{thinshell}:

\begin{align}
	\mathcal{H} &= \sqrt{h}\; \rho(r_s)\; \delta(r-r_s)\; \sqrt{g^{rr}p_r^2 + m_0^2},\label{eq:hamiltonian17}\\
    \mathcal{H}_a &= \delta^r_a \; \sqrt{h}\; \rho(r_s)\;\delta(r-r_s) \; p_r,
\end{align}

where $\rho(r_s)$ is a scalar function yet to be determined and $h_{ij}$ is the metric on $S^2$ induced by $g_{ij}$. Imposing the condition that in this process the number of particles is constant, gives us \cite{thinshell}:

\begin{equation}
	\rho(r_s) \int d\theta d\phi \sqrt{h(r_s)} = 4\pi n.\label{eq:rho}
\end{equation}

If we rescale the momentum and mass as $ p_s=np_r $ and $M = nm_0$, the constraints (\ref{eq:const1}), (\ref{eq:const2}) and (\ref{eq:const3}) become \cite{tutorial}:

\begin{align}
	& -\frac{1}{6\sigma\mu^2}\Big[\sigma^2 \mu s^2 + 4 f^2 \mu^3 - 4 f \sigma \mu^2 s + 12 \sigma \mu \sigma''\nonumber\\
    &\quad -12 \sigma \sigma' \mu' - 3\mu (\sigma')^2 - 12\sigma \mu^3 - \expect{p}^2 \sigma^2\mu^3\Big]\nonumber\\
    &= \delta(r-r_s) \sqrt{p_s^2/\mu^2 + M^2}\label{eq:hm},\\
    &\mu f' - \frac{1}{2}s\sigma' = -\frac{p_s}{2}\delta(r-r_s)\label{eq:diffeom},\\
    & \mu f + s\sigma - \expect{p}\mu\sigma = 0.\label{eqn:cmcm}
\end{align}

CMC constraint (\ref{eqn:cmcm}) gives us $s = \expect{p}\mu - \mu f /\sigma$ and when put in (\ref{eq:diffeom}) yields \cite{tutorial}:

\begin{equation}
	\frac{\mu}{\sqrt{\sigma}}\left(f\sqrt{\sigma}-\frac{1}{3}\expect{p}\sigma^{3/2}\right)' = -\frac{p_s}{2}\delta(r-r_s)\label{eq:diffeomdelta}.
\end{equation}

A method to solve (\ref{eq:diffeomdelta}) is given in \cite{thinshell,tutorial}, the solution is:

\begin{equation}
	f(r) = \frac{1}{3}\expect{p}\sqrt{\sigma} + \frac{A_-}{\sqrt{\sigma}}\Theta(r_s-r) + \frac{A_+}{\sqrt{\sigma}}\Theta(r-r_s)\label{eq:f},
\end{equation}

where $\Theta$ is the Heaviside distribution and

\begin{equation}
	A_+ - A_- = -\frac{p_s \sqrt{\sigma(r_s)}}{2\mu(r_s)}.\label{eq:aa}
\end{equation}

Here $A_-$ and $A_+$ are functions of $r_s$ and $p_s$. On the other hand, if we look at (\ref{eq:hm}), the only singular part on the left hand side comes from the $-2\sigma''/\mu$ and we have \cite{tutorial}:

\begin{equation}
	\text{sing}\left(\frac{2\sigma''}{\mu}\right) = -\delta(r-r_s) \sqrt{p_s^2/\mu^2 + M^2},
\end{equation}

which yields

\begin{equation}
	\text{sing}(\sigma'') = -\frac{1}{2}\delta(r-r_s) \sqrt{p_s^2 + M^2 \mu^2}.
\end{equation}

Let $\gamma = \lim_{r\to r_s^+}\sigma'$ and $\kappa = \lim_{r\to r_s^-}\sigma'$, then we have \cite{tutorial}:

\begin{equation}
	\gamma - \kappa = -\frac{1}{2} \sqrt{p_s^2 + M^2 \mu^2(r_s)}.\label{eq:gammaminuskappa}
\end{equation}

Last but not least, we need to impose continuity to $\mu$. Using (\ref{eq:mu2}) we obtain:

\begin{multline}
	\kappa^2\Big(A_-^2\sigma^{-1}(r_s) + (\frac{2}{3}\expect{p}A_- - 8m_-)\sigma^{1/2}(r_s)\\
     +4\sigma(r_s) + \frac{1}{9}\expect{p}^2\sigma^2(r_s)\Big)^{-1} = \gamma^2 \Big(A_+^2\sigma^{-1}(r_s) \\
     + (\frac{2}{3}\expect{p}A_+ - 8m_+)\sigma^{1/2}(r_s)
     +4\sigma(r_s) + \frac{1}{9}\expect{p}^2\sigma^2(r_s)\Big)^{-1}
\end{multline}

In the next section we investigate the dynamics of the spherically symmetric thin shell of dust.

\section{Evolution of the shell}\label{sec:evolution}

The presymplectic form is as follows \cite{thinshell}:

\begin{equation}
	\theta = \int dr d\theta d\phi p^{ab}\delta g_{ab} + 4\pi p_s\delta r_s \label{eq:presymp}.
\end{equation}

When definitions of metric and conjugate momentum of metric which are defined in (\ref{eq:gab}) and (\ref{eq:pab}) are taken into account in equation~\ref{eq:presymp} and the angular part is integrated out, what we have is the following:

\begin{equation}
	\theta = 4\pi \int_{0}^{1} dr (2f\delta \mu + s\delta\sigma) + 4\pi p_s\delta r_s.
\end{equation}

Using the CMC condition (\ref{eqn:cmcm}) we find (modulo an exact form):

\begin{equation}
	\theta = 4\pi \int_0^1 dr \left(\expect{p}\mu\delta\sigma - \frac{2\mu}{\sqrt{\sigma}}\delta(f\sqrt\sigma)\right) + 4\pi p_s\delta r_s.
\end{equation}

Let us use the form of $f$ found in (\ref{eq:f}), then we obtain:

\begin{multline}
	\theta = \underbrace{4\pi \int_0^1 dr \left[\expect{p}\mu\left(1-\frac{2}{3\sqrt{\sigma}}\right)\delta\sigma - \frac{2\mu\sqrt{\sigma}\delta\expect{p}}{3}\right]}_{\equiv -\phi} \\ - 8\pi\left[\delta A_- \int_0^{r_s} dr \frac{\mu}{\sqrt{\sigma}} + \delta A_+ \int_{r_s}^1 dr \frac{\mu}{\sqrt{\sigma}}\right].
\end{multline}

Here we move on to the isotropic coordinates: $\mu = \sqrt{\sigma}/r$. In this case, the presymplectic potential becomes $\theta = -\phi + 8\pi\; \delta(A_+ - A_-)\ln r_s$ modulo an exact form. Using equation (\ref{eq:aa}) we can write this as follows:

\begin{equation}
	\theta = -\phi - 4\pi\; \delta(p_s r_s) \ln r_s.
\end{equation}

The symplectic form, $\omega$, which is minus the exterior derivative of $\theta$ is found out to be:

\begin{equation}
	\omega = -\delta\theta = \delta\phi + 4\pi\;\delta r_s \wedge \delta p_s,
\end{equation}

where $\delta \phi$ does not depend on $\delta r_s$ or $\delta p_s$. In matrix form, $\omega$ can be represented as follows:

\begin{equation}
	\omega_{ab} = \begin{pmatrix}
		F(\delta\phi) &0 &0 \\
                0 & 0 & 4\pi\\
                 0&-4\pi & 0
	\end{pmatrix},
\end{equation}

where the order of the coordinates is as $\text{``others"},r_s,p_s$ where $F$ is some function of $\delta\phi$. The inverse of the symplectic form to calculate the Poisson  brackets is found as follows:

\begin{equation}
	\omega^{ab} = \begin{pmatrix}
		F^{-1}(\delta\phi) &0 &0 \\
                 0& 0 & -1/4\pi\\
                 0&1/4\pi & 0
	\end{pmatrix}.
\end{equation}

$\mathcal{H}$ is the rescaled Hamiltonian density in equation~(\ref{eq:hm}). Using equation~(\ref{eq:hamiltonian17}) and equation~(\ref{eq:rho}), the Hamiltonian is then found out to be the following:

\begin{align}
	H &= \int dr d\theta d\phi \sqrt{g} \mathcal{H}\nonumber\\
      &= \pi^2 \sigma(r_s) \sqrt{p_s^2 + M^2 \mu^2(r_s)}. \label{eq:h2}
\end{align}

Now, we find the equations of motion for $r_s$ and $p_s$. We have already calculated the symplectic form so we can calculate the Poisson brackets of $r_s$ and $p_s$ with the Hamiltonian. For $\dot{r}_s$ we have:

\begin{align}
	\dot{r}_s &= \{r_s,H\},\nonumber\\
    &= \frac{1}{4\pi}\frac{\partial H}{\partial p_s},\nonumber\\
    &= \frac{\pi}{4}\sigma(r_s) \frac{p_s}{\sqrt{p_s^2 + M^2 \mu^2(r_s)}},\nonumber\\
    &= \frac{\pi}{4} \sigma(r_s) \frac{p_s}{\sqrt{p_s^2 + M^2 \sigma(r_s)/r_s^2}}. \label{eq:rst}
\end{align}

And for $\dot{p}_s$ we have the following:

\begin{align}
	\dot{p}_s &= \{p_s,H\},\nonumber\\
    &= -\frac{1}{4\pi} \frac{\partial H}{\partial r_s},\nonumber\\
    &= -\frac{\pi}{4}\sigma'(r_s)\sqrt{p_s^2 + M^2 \mu^2(r_s)} \nonumber\\
    &\quad - \frac{\pi}{4}\sigma(r_s)\frac{M^2 \mu(r_s)\mu'(r_s)}{\sqrt{p_s^2 + M^2 \mu^2(r_s)}},\nonumber\\
    &= -\frac{\pi}{4}\sigma'(r_s)\sqrt{p_s^2 + M^2 \sigma(r_s)/r_s^2} \nonumber\\
    &\quad - \frac{\pi}{4}\frac{\sigma(r_s)}{2 r_s^2}\frac{M^2}{\sqrt{p_s^2 + M^2 \sigma(r_s)/r_s^2}}\nonumber\\
    &\quad \quad \times \left(\sigma'(r_s)-\frac{2\sigma(r_s)}{r_s}\right).\label{eq:pst}
\end{align}

These equations are very complex to be solved analytically. We will make an approximation and suppose that the shell is very close to origin: $r_s \ll 1$.

\subsection{Perturbation Analysis}

Near the origin the metric should be flat and for that purpose $\sigma(r)\approx r^2$. There is one more issue to be solved: $\sigma'(r)$ is discontinuous at $r = r_s$. Therefore we will equate $\sigma'(r_s)$ to the average of left ($\kappa$) and right ($\gamma$) limits of $\sigma'(r_s)$. Since we can approximate $\sigma(r)\approx r^2$ inside $r_s$, we find $\kappa = 2 r_s$. The use of equation (\ref{eq:gammaminuskappa}) yields:

\begin{equation}
	\sigma'(r_s) = \frac{\kappa+\gamma}{2} \approx -\frac{1}{4}\sqrt{p_s^2+M^2} + 2r_s.\label{eq:sigmaprimers}
\end{equation}

For $r_s \ll 1$, the equations of motion (\ref{eq:rst}) and (\ref{eq:pst}) become:

\begin{align}
	\dot{r}_s &\approx \frac{\pi}{4}\frac{r_s^2 p_s}{\sqrt{p_s^2 + M^2}},\label{eq:rsapprox}\\
    \dot{p}_s &\approx \frac{\pi}{16}(p_s^2 + M^2) - \frac{\pi}{2}r_s\sqrt{p_s^2 + M^2} + \frac{\pi M^2}{32} \label{eq:psapprox}.
\end{align}

\begin{figure}
	\includegraphics[width=0.4\textwidth]{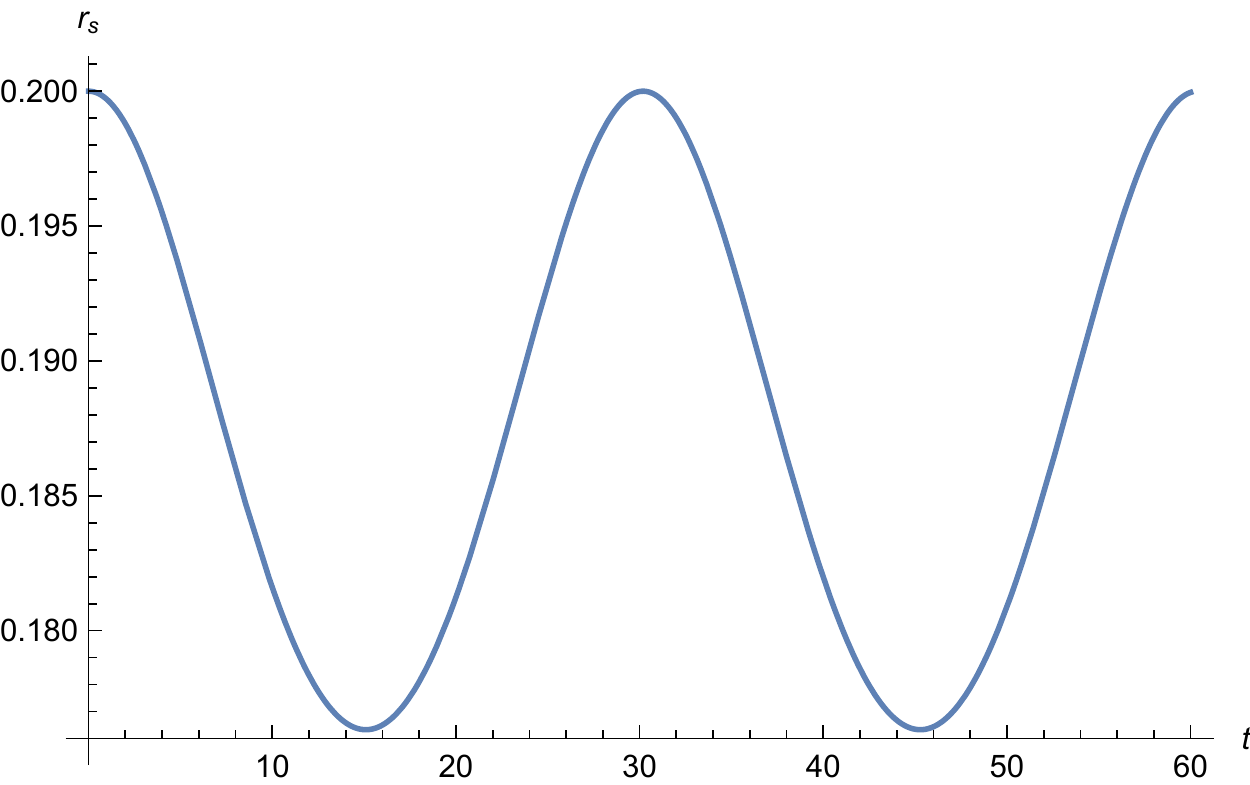}
    \caption{The evolution of $r_s$ for $r_s(0)=0.2$, $p_s(0)=0$ and $M=1$.}
    \label{fig:rs02}
\end{figure}

\begin{figure}
	\includegraphics[width=0.4\textwidth]{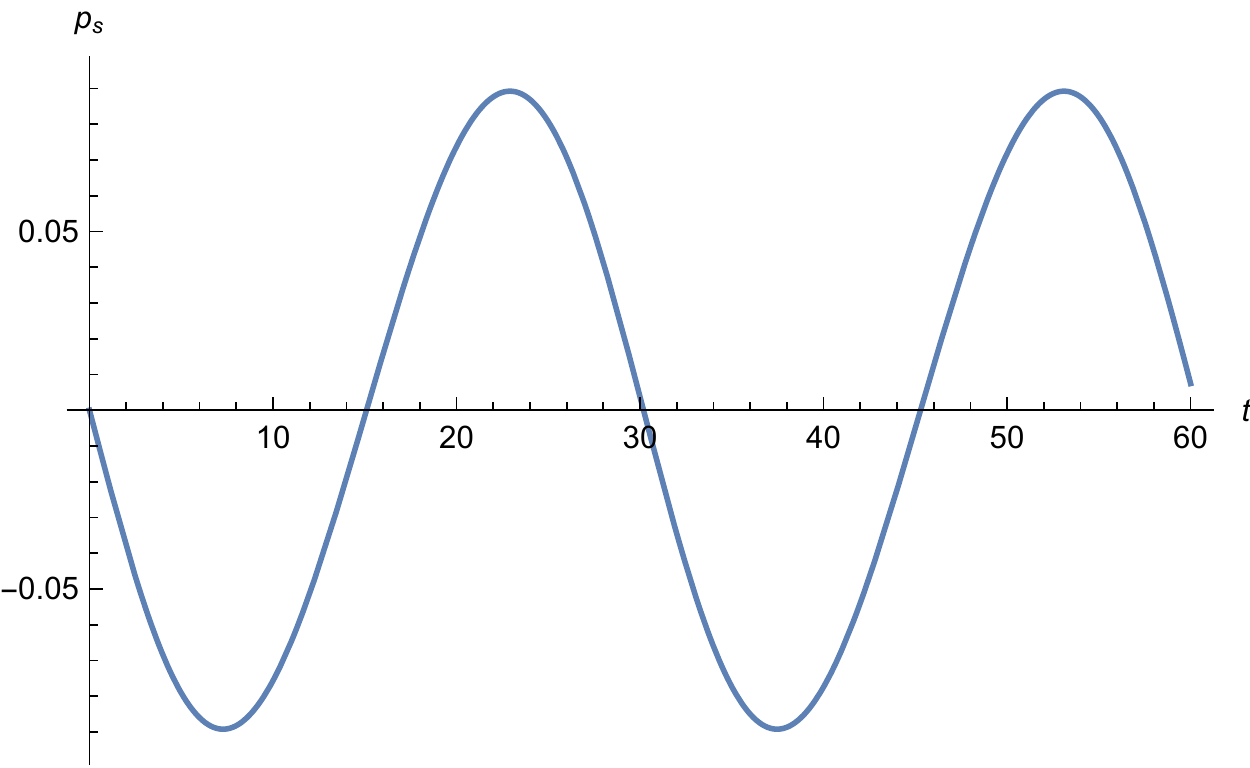}
    \caption{The evolution of $p_s$ for $r_s(0)=0.2$, $p_s(0)=0$ and $M=1$.}
    \label{fig:ps0}
\end{figure}

These equations admit a static solution: $r_s = 3M/16,\; p_s = 0$. If we do a perturbation analysis around the static solution the results turn out to be functions of circular functions:

\begin{align}
	r_s(t) &\approx \frac{3M}{16} + A_r \sin\left(\Omega(t-t_r)\right),\\
    p_s(t) &\approx A_p \sin\left(\Omega(t-t_p)\right),
\end{align}

where $\Omega = 3\pi M^{3/2}/32\sqrt{2}$, $A_r,A_p,t_r,t_p$ are integration constants and there are conditions on the first two integration constants: $|A_r| \ll 3M/16 \ll 1, |A_p| \ll M$. As we see in this limit $r_{s}$ is oscillating and it never reaches $r=0$ point. If a black hole was to form, $r_s$ would always be decreasing unlike the current result. Therefore in this approximation black holes do not form. See Figure~\ref{fig:rs02} and Figure~\ref{fig:ps0}.

\subsection{High Momentum Limit: $-p_s \gg M$}

In the limit $-p_s \gg M$, the equations of motion become the following two:

\begin{align}
	\dot{r}_s &\approx -\frac{\pi r_s^2}{4} ,\\
    \dot{p}_s &\approx \frac{\pi p_s^2}{16}.
\end{align}

The solutions are given as follows:

\begin{align}
	r_s(t) &\approx \frac{1}{1/r_s(0) + \pi t/4},\\
    p_s(t) &\approx \frac{1}{1/p_s(0) -\pi t/16}.
\end{align}

In the high momentum limit, the dust shell collapses to a single point in the $t \to \infty$ limit, however it does not reach the origin in a finite amount of time. More importantly, the shell does not enter a parallel universe. In ref. \cite{birkhoff} it is found out that the spherically symmetric vacuum solution of SD in an asymptotically flat space is an Einstein-Rosen bridge: a portion of the usual Schwarzschild solution in isotropic coordinates. Now, we did not find that this result will be a product of gravitational collapse: just like maximally extended Schwarzschild solution is not obtained by a gravitational collapse. What is interesting in our solution is that the radial momentum of the shell keeps increasing (in magnitude it decreases) and when it becomes comparable with minus the mass of the shell, $-M$, this limiting analysis is no longer valid. Therefore there is still the possibility that a singularity might not form. More detailed future studies are needed to remedy this shortcoming.

\section{Conclusion}\label{sec:conclusion}

In this study we investigated the gravitational collapse of a shell of dust in shape dynamics (SD). SD is a theory of gravity that is equivalent to general relativity (GR) in the constant mean extrinsic curvature gauge of the latter. In SD, instead of the local Lorentz symmetry of GR there is local scale (Weyl) symmetry. We considered a filled-in ball of radius one, with antipodal matching at the surface, as our space.

We obtained the full equations of motion for the radial coordinate radius, $r_s$, of the shell and its total radial momentum, $p_s$. We solved the equations in the approximation where the shell is close to the origin: $r_s\approx 0$. We have found out a static solution for $r_s = 3M/16 \ll 1$. Moreover we did a perturbation analysis around this static solution and found out that $r_s$ and $p_s$ are periodic with the angular frequency $3M/8\sqrt{2}$.

We also studied the large momentum limit ($-p_s \gg M$) of the equations in order to see whether a large infalling radial momentum will break the periodicity of the previous equations. The result is that $\lim_{t\to\infty} r_s = 0$. This is in contrast to shells reaching the singularity in a finite amount of time in GR. However this analysis will break down when $-p_s$ becomes comparable with the mass of the shell, $M$. Therefore it is still an open question whether a singularity will form or not. We hope that more detailed future works may be able to answer this point.

\section{Acknowledgements}

Authors are grateful to Edward Anderson, Teoman Turgut, Mikhail Sheftel, Soley Ersoy, Erol Barut, Deniz Yılmaz and Medine İldeş for useful discussions. F.S.D. is supported by TUBİTAK 2211 Scholarship. The research of F.S.D. and M.A. is partly supported by the research grant from Boğaziçi University Scientific Research Fund (BAP), research project No. 11643.

\bibliographystyle{unsrt}
\bibliography{refs}

\end{document}